\documentclass[%
 reprint,
 amsmath,amssymb,
 aps,
]{revtex4-2}
\usepackage{graphicx}
\usepackage{booktabs}
\usepackage{dcolumn}
\usepackage{xr}
\usepackage{bm}
\usepackage{hyperref}
\hypersetup{
    colorlinks = true,
    citecolor = {blue},
    urlcolor = {blue},
    linkcolor = {blue}
}
\usepackage{chemformula}
\usepackage{lipsum}
\usepackage{color}

\newcommand{\aza}{C$_{59}$N}
\newcommand{\azam}{C$_{59}$N$^-$}
\newcommand{\azar}{C$_{59}$N$^\bullet$}
\newcommand{\NV}{NV$^-$}
\newcommand{\nvz}{NV$^0$}
\newcommand{\Ns}{N$_{\rm s}$}
\newcommand{\aun}{$a_1$}
\newcommand{\ex}{$e_x$}
\newcommand{\ey}{$e_y$}

\begin{document}

\title{Interaction between shallow \NV \ and spin active azafullerenes\\ on hydrogenated and fluorinated (001) diamond surfaces}

\author{Bastien Anézo}
\email{bastien.anezo@cnrs-imn.fr}
\affiliation{Nantes Université, CNRS, Institut des Matériaux de Nantes Jean Rouxel, IMN, F-44000 Nantes, France}
\affiliation{Jožef Stefan Institute, Jamova 39, SI-1000 Ljubljana, Slovenia}

\author{Denis Arčon}
\affiliation{Jožef Stefan Institute, Jamova 39, SI-1000 Ljubljana, Slovenia}
\affiliation{Faculty of mathematics and physics, University of Ljubljana,  Jadranska 19, SI-1000 Ljubljana, Slovenia}
 
\author{Chris Ewels}
 \email{chris.ewels@cnrs-imn.fr}
\affiliation{
Nantes Université, CNRS, Institut des Matériaux de Nantes Jean Rouxel, IMN, F-44000 Nantes, France}

\date{\today}

\begin{abstract}

The interaction between surface-lying nitrogen-substituted fullerenes (radical azafullerene, \azar) with sub-surface negative nitrogen-vacancy complexes (\NV) in diamond is investigated using first principles calculations. We consider (2$\times$1) reconstructed (001) oriented diamond surfaces with both H- and F-surface termination.
The charge stability of \NV \, when in close proximity to both the nearby surface and the spin active azafullerene is discussed, in the context of diamond band bending arising from surface-induced changes in electron affinity (EA).
In the case of the hydrogenated surface, the system spin is quenched, yielding a negatively charged azafullerene (\azam) and neutrally charged NV$^0$ as the most stable electronic configuration.
In contrast, fluorinating the surface favours the negatively charged \NV\, and conserves the \azar\, neutrality and stabilizes uncompensated free spins.
This opposing behaviour is attributed to surface charge doping emerging from different band bending effects associated with the surface EA.
This study is consistent with experimentally observed photoluminescence quenching, and shows that surface passivation by fluorination could efficiently tackle the problem of charge transfer between adsorbed molecules and shallow NV centers.
\end{abstract}

\maketitle

\section{\label{sec:level1}Introduction}

Nanomaterials and molecular systems have sucessfully achieved important milestones in magnetism  and quantum information processes (QIP) in recent years \cite{Coronado_magnetism_2020, Gaita-arino_molecular_QC_2019}.
Semiconducting carbon based nanomaterials such as epi-graphene, nanotubes and fullerenes are of interest for QIP, since they are capable of exhibiting both localised long-lived coherent spin states, as well as high-mobility conductivity \cite{SWN_spin_chen_long-lived_2023, Epigraphene_Zhao_ultrahigh-mobility_2024}.
Notable amongst these are endohedral fullerenes, carbon cages containing trapped species within them, such as $\rm N@C_{60}$.
The nuclear spin states of the trapped nitrogen atom can serve as storage for quantum information, and a delocalised electron over the cage as a read/write probe \cite{Kane_silicon-based_1998}.
$\rm N@C_{60}$ shows competitive longitudinal (T$_1$) and dephasing (T$_2$) nuclear spin decoherence times already at room temperature, satisfying one of the important criteria for effective QIP candidates \cite{Meyer2003, Divincenzo_physical_2000}. However, its low production yield and complicated purification procedures are the two main limiting factors that hinder further development in that direction.  

\begin{figure}[!t]
\includegraphics[scale=0.35]{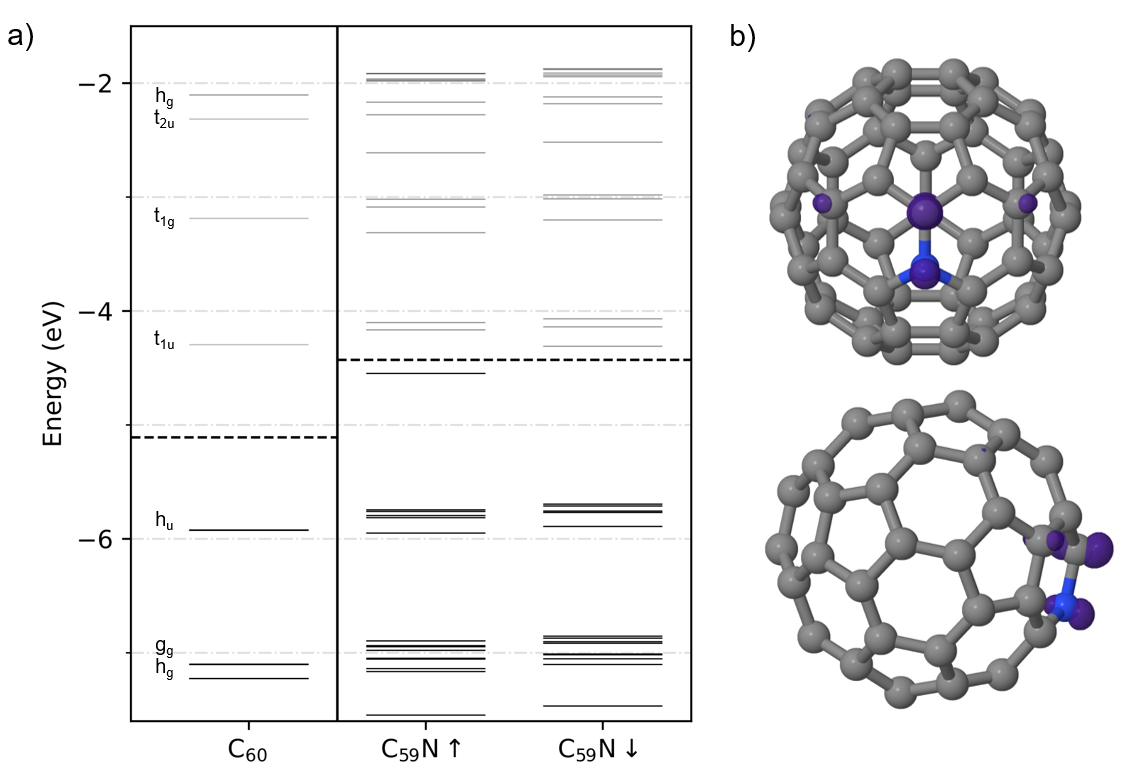}
\caption{The energies in a) represent C$_{60}$ (left) and \azar\, (right) eigenstates with symmetries and spin orientation. Zero is set to the vacuum level of each molecule. Dashed lines correspond to respective Fermi levels. Full and faded bars show occupied and unoccupied states respectively. With b) the \azar\, radical state wave function is shown: the visualisation has been done using Jmol. Grey and blue spheres represent carbon and nitrogen atoms, respectively,\, while purple lobes show the wave function isosurface with a $7.5$~meV.$\mathring{\rm A}^{-3}$ cut-off.}
\label{fig:figure.aza}
\end{figure}

An alternative fullerene platform is the paramagnetic azafullerene \aza, where substitution of a carbon by a nitrogen atom in Buckminsterfullerene, C$_{60}$, breaks a C--C $\pi$ bond. This leaves a free valence electron which is predominantly localised in the vicinity of the C--N bond (Fig.\ref{fig:figure.aza}.b).
This symmetry breaking splits the degenerate states of C$_{60}$ (h$_g$, g$_g$, h$_u$ etc ...) into a spread of eigenstates energetically close to Buckminsterfullerene associated states. In particular, the triply degenerated t$_{1u}$ lowest unoccupied molecular orbital (LUMO) state splits, trapping the radical electron and creating a semi-occupied molecular orbital (SOMO).
Electron paramagnetic resonance (EPR) investigations of this SOMO, through photolysis in a CS$_2$ solvated liquid phase or by thermolysis in its dried solid state, corroborate the unpaired spin localization next to the N-site and  report outstanding $T_1$, $T_2$ and electronic spin lifetime \cite{Photo_light-induced_1997,Photolysis_Gruss_1997,thermolysis_Ferenc_1999, arcon2007stability,  Robust_tanuma_robust_2021,Long_tanuma_2023}.
These properties are a good starting point for any application, but effective spin control of such molecular systems requires good control over the system geometry in either one- (1D) or two-dimensions (2D).
Recent studies have shown how to engineer an assembly of such molecular radical entities by depositing \azar \, directly onto an (111)Au surface or a [10]CPP molecular monolayer template, while still conserving the spin active character of the system \cite{Noncontact_tanuma_2023, kladnik_engineering_2025}.

To read-out locally the single molecular spin properties we could use a negatively charged nitrogen-vacancy (\NV) defect in diamond. This impurity has been successfully used as a nanometric probe in magnetometry and biosensing as well as for single-photon generation for QIP
\cite{Rondin_magnetometry_2014,Janitz_diamond_2022,Jacques_experimental_2007}.
\NV \, consists of a substitutional nitrogen (\Ns) impurity neighbouring a carbon vacancy in the diamond lattice  and an additional captured electron.
Typically \Ns \, concentration control is performed through ion implementation and the vacancies are made mobile with an annealing process to create a stable NV defect structure.
Electronically, the centre consists of a \Ns \, lone pair, the three  electrons originating from the unpaired carbon orbitals neighbouring the vancancy, and an electron captured from a nearby substitutional donor (usually P or N).
This creates four electronic states (\{$a'_1$,\aun,\ex,\ey\}) containing six electrons. The $a_1'$ orbital lies in the diamond valence band, while the four remaining electrons populate the singlet $a_1$ and doublet $e_x$, $e_y$ gap states.
In the most stable $^3$A$_2$ symmetry, the defect has a total spin of $S=1$.
These highly localised states lying deep inside the wide diamond gap are well decoupled from the host lattice and can be treated  as an isolated system.  This enables the use of fluorescence to access the NV$^-$ electron spin resonance (ESR) and to optically detected magnetic resonance (ODMR) response.
Bringing this defect closer to a source of magnetic noise (spin active systems), one can perform magnetometry and/or relaxometry with high sensitivity and spatial resolution. Promising experimental studies have demonstrated that NV$^-$ centres are indeed sensitive to surface spins \cite{Gado_steinert_magnetic_2013, TEMPO_grotz_sensing_2011}.
Notably, Pinto \textit{et al} have successfully shown that the ODMR of \NV\, in bulk diamond is capable of detecting multi-layer surface deposited $\rm N@C_{60}$ \cite{D.Pinto_readout_2020}.
This result provides a working proof of principle that \NV \, can indeed act as a read-write atomic-size element for fullerene-based QIP.

However, surface effects are important when performing measurements with shallow \NV\,  defects. The main property at play is the electron affinity (EA), it describes how much a system stabilises over an electron gain, here defined as the difference between the vacuum level and the conduction band minimum of the diamond. It is expressed through an electronic band banding effect at the surface interface. This has been extensively studied for both H- and F-terminated surfaces for a large variety of surface orientations and reconstructions \cite{Tiwari_PhysRevB.84.245305,Robertson_robertson_band_1998,kaviani_proper_2014,DIEDERICH_1998219,Bandis_PhysRevB.52.12056,BANDIS_1996315,Rietwyk_10.1063/1.4793999,Humphreys_10.1063/1.118545,Rutter_PhysRevB.57.9241,BAUMANN_1998320,Cui_PhysRevLett.81.429,Kern_PhysRevB.56.4203,LI_2019273}. The electron affinity is known to influence the interaction of the diamond with on-surface molecules, as it may promote charge transfer between the substrate and the adsorbent. In the case of NV charge states, it translates as a conversion between \NV\, and \nvz\, where only the former can be utilized in NV magnetometry measurements. As a possible mitigation approach to avoid 
charge quenching of shallow \NV\,  centers, wetting of the diamond surface has been suggested \cite{Neethirajan_doi:10.1021/acs.nanolett.2c04733}.

Probing the spin state of \azar\rm, adsorbed on a diamond surface, requires two main conditions to be fulfilled: (1) the use of shallow \NV\, centers to achieve suitable sensitivity while (2) maintaining  the \NV\,  charge integrity intact. Because of the high electron affinity of fullerenes, it is not easy to satisfy both conditions at the same time. Moreover, the wetting of the diamond surface approach \cite{Neethirajan_doi:10.1021/acs.nanolett.2c04733} is not applicable in such cases as fullerenes are poorly soluble in water.

In this DFT study we investigate the coupling of  \azar\, to shallow \NV\, centres when the former are adsorbed on the surface of nanodiamonds.
Such calculations are usually computationally prohibitive due to their complexity and low symmetry, requiring large cell sizes.
In the current study we circumvent this problem by using smaller wave function basis sets to describe (001) oriented diamond surfaces in 'slab' calculations.  Notably, we model H- and F-terminated surfaces and explore the electronic stability of the surface deposited spin active molecule \azar, considering a shallow \NV\,  below. 

\section{\label{sec:level2}Method}

Density Functional Theory (DFT) investigations were carried out using the AIMPRO package, a localised orbital electronic structure code \cite{AIMPRO_1_jones_chapter_1998,AIMPRO_2_rayson_rapid_2008,AIMPRO_3_rayson_highly_2009}.
Exchange-correlation was described using the generalised-gradient approximation (GGA) via the PBE functional \cite{GGA_PhysRevLett.77.3865} coupled with D2-Grimme long-range dispersion corrections \cite{Grimme_1}.
Relativistic Hartwigsen-Goedecker-Hutter (HGH) pseudopotentials  were used to describe the core electrons \cite{HGH_PhysRevB.58.3641}.
Brillouin zone sampling was handled with the Monkhorst-Pack scheme using $n\times n\times n$ or $n\times n\times 1$ k-point grids \cite{Monkhorst_special_1976}. \\
A similar approach has been previously and successfully applied to model both \azar \cite{Robust_tanuma_robust_2021,Long_tanuma_2023,Noncontact_tanuma_2023}  and \NV\, in diamond \cite{H.Pinto_bandst_PhysRevB.86.045313,Meara_positive_PhysRevB.100.104108,Sque_structure_2006}.The plane-wave energy cut-off for charge density was set to 175 Ha in all calculations. A Fermi distribution function is used to fill the levels to help with convergence, with an effective temperature of the electronic system of  k$_B$T = 0.025~eV.  \\

Calculations were considered self-consistent and converged when the self-consistent energy difference, total energy and position change between two iterations was less than 10$^{-6}$~Ha, 10$^{-6}$~Ha and 10$^{-5}$~a.u respectively. \\

\begin{figure}[!t]
\includegraphics[scale=0.37]{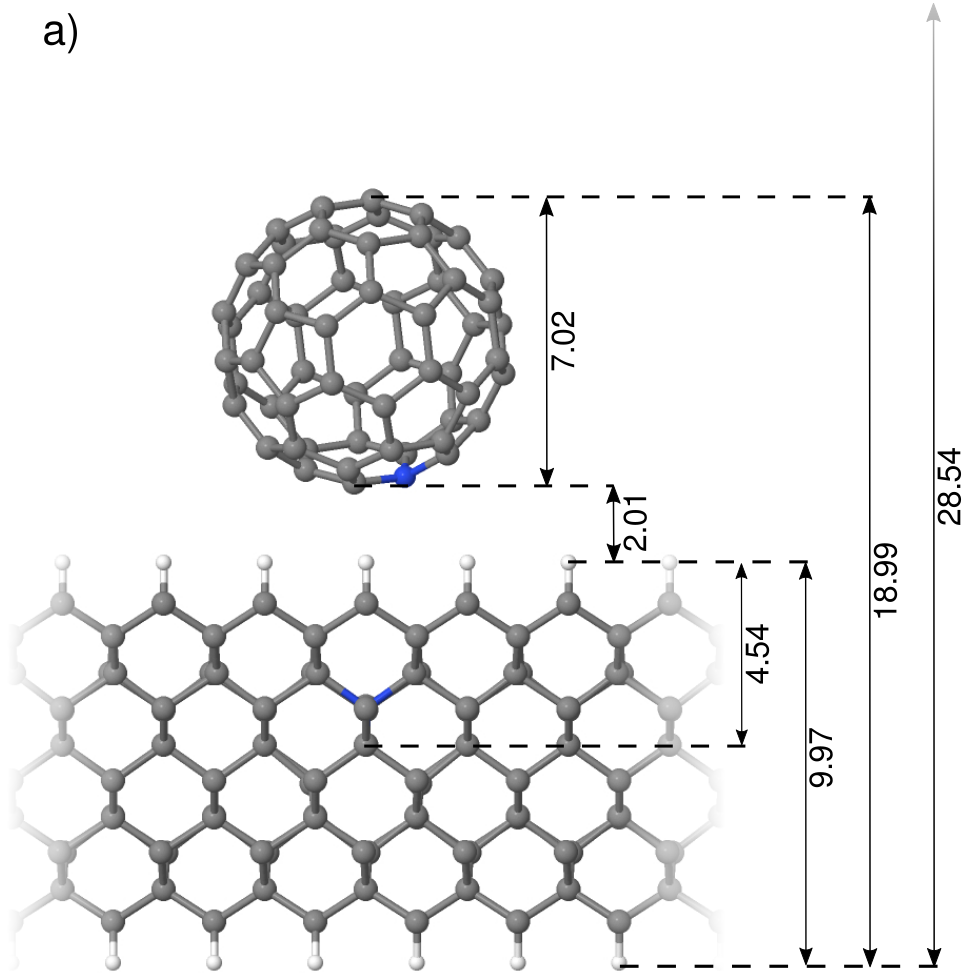}
\par\vspace{0.5cm}
\includegraphics[scale=0.30]{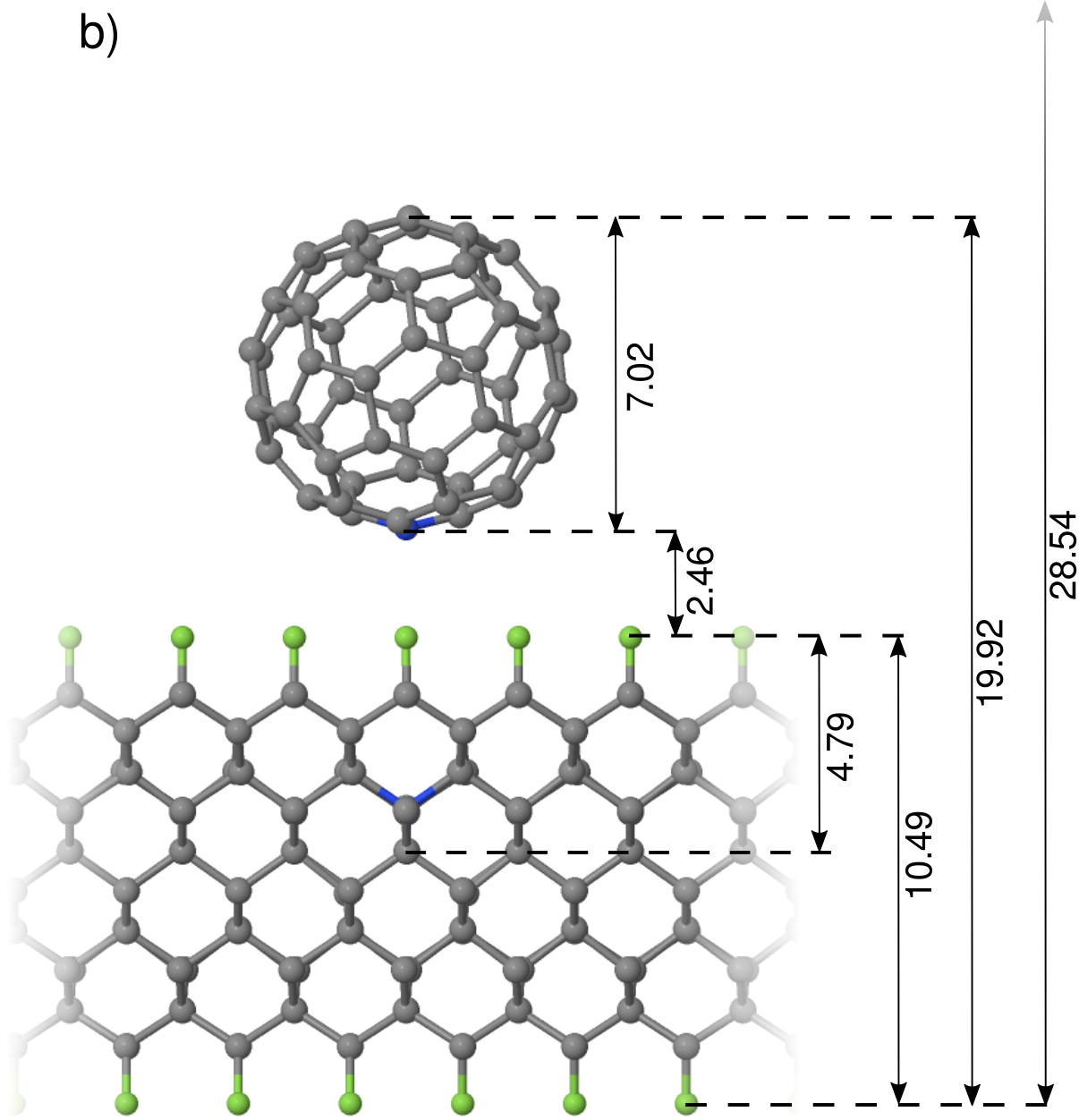}
\caption{ Hetero-system geometries after complete structural optimisation for a) hydrogenated and b) fluorinated surface. Relevant c-axis lengths (\AA): fullerene diameter and distance to the surface, slab thickness, NV defect distance to surface, total system width and c-axis unit cell length (including vacuum).}
\label{fig:figure.structure}
\end{figure}

\begin{figure}[!t]
\includegraphics[scale=0.47]{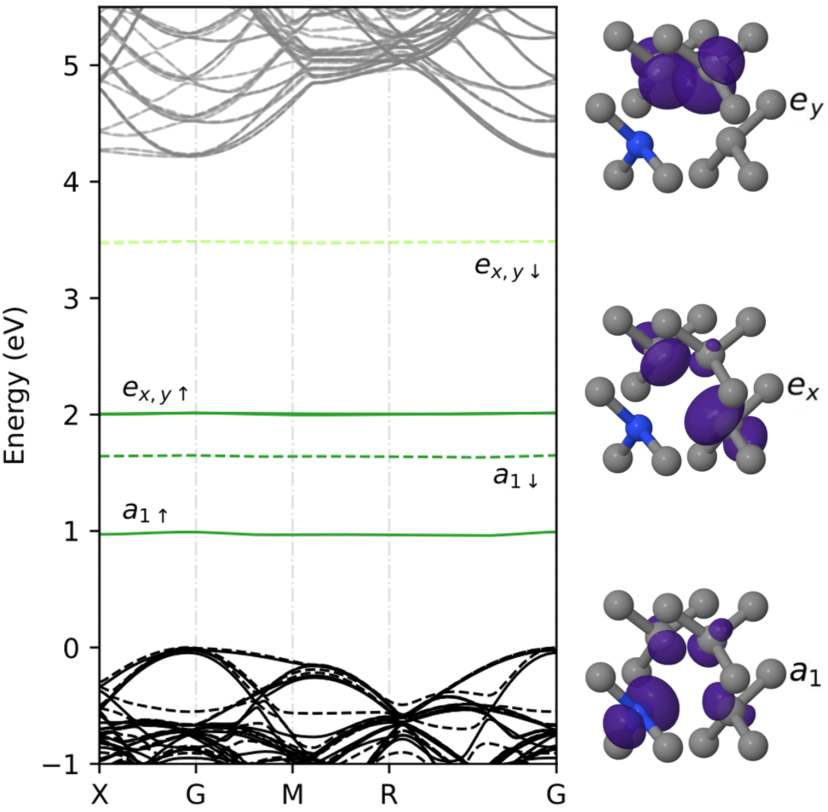}
\caption{Electronic band structure of \NV\,  within a 4$\times$4$\times$4 diamond supercell. Zero is set to the top of the diamond valence band. Solid/dashed lines represent spin-up/-down electrons, respectively, while dark/faded lines showing filled/empty states, respectively. Black and green colours indicate bulk diamond and NV defect related eigen states. The wavefunction isosurfaces of the \aun\,  and $e$ states are shown  with a 20~meV$.\mathring{\rm A}^{-3}$ cut-off.}
\label{fig:figure.supercell}
\end{figure}

Following this scheme, isolated \azar\, was geometrically converged in a large cell (a = c = 15.9~$\mathring{A}$) to avoid interaction between molecules in neighbouring cells (eigenfrequencies were checked to ensure the minimum in energy was reached).  In the case of 3D bulklike supercells, all atoms were allowed to move in the geometric optimisation before any further analysis.
For 2D slabs, all atoms were allowed to move to optimise the slab structure. After this all atoms were fixed when \NV was introduced,  except those directly associated with the defect i.e. \Ns\, and the vacancy neighbours. 
For the hetero-systems with an azafullerene on the diamond surface, all atoms were fixed except the NV and \azar\, related atoms.
Multiple initial fullerene positions and orientations were optimised to find the overall global minimum.

Fig.\ref{fig:figure.structure} shows the modeled slab unit cells (383 atoms). They measure 1.4~nm in the x and y directions and are about 1~nm thick, with a 1.8~nm vacuum to ensure no interaction between the top slab surface (or the \aza\, molecule) and the bottom surface of the next repeating slab. In the combined structure (443 atoms) the NV lies 0.45~nm (0.47~nm) from the surface and 0.65~nm (0.73~nm) from the azafullerene base for the hydrogenated (fluorinated) surfaces respectively.

To reduce the computational cost of such large systems we describe the diamond slab C atoms with a wave function basis set of 22 Gaussian-based function of s, p and d character. For the fullerene we use a larger 38 function set due to its higher degrees of freedom and the vacuum proximity. This basis set has previously been demonstrated to be reliable\cite{Robust_tanuma_robust_2021,Long_tanuma_2023}

For periodic systems it is necessary to maintain overall charge neutrality in the cell. Therefore in order to model a negatively charged NV defect, an equal and opposite uniform constant positive electric potential is applied throughout the cell.  This approach is used for all negatively charged bulk and surface calculations, also when modeling \aza, on both H- and F-terminated surfaces. No charge correction is applied on the charged systems, thus we take the relevant valence band maximum as reference to compare their energies. Mulliken analysis was used to generate density of states projected (pDoS) onto specific atoms or atom groups.

Rotating the \aza, from pointing to the diamond surface to facing the vacuum, we found the configuration with the radical pointing towards the surface to be the most stable orientation. We then varied the \azar\, position above a H or F atom and two different hollow sites.   The total system energy varied by only a few 10meV, while not affecting the electronic configuration regardless of the choice of site.

\subsection{Discussion concerning choice of functional} \label{fucntional}

We note that the HSE06 exchange-correlation functional has been found to capture electronic properties of \NV\, defects in bulk diamond with high accuracy, namely the electronic band gap, zero phonon line (ZPL) and ionization energy \cite{deak_accurate_2011,gali_theory_2009}. The approach has also been successfully used to study diamond surfaces and the impact of different orientations and terminations on the \NV gap states \cite{shen_epoxy_2021, kaviani_proper_2014,chou_nitrogen-terminated_2017}. In contrast, the GGA-PBE method used here gives qualitative accuracy of the same physical values (ZPL) and can show sigificant discrepancy for properties such as band gap, and may therefore not seem an obvious choice. There are however a number of reasons GGA-PBE is adopted here.\\
The first is the prohibitive computational cost of HSE06 (typically from 2 to 10 times slower than GGA-PBE depending on the basis set type) which places such calculations beyond our available computational resources (the calculations already require a reduced basis set size as mentionned below).  The second concerns the choice of mixing and screening cut-off parameters within HSE which can be system-dependent. In general for bulk systems the HSE06 functional works well. However precautions have to be taken when describing systems with low dielectric constant, large surfaces, as well as for layered materials \cite{deak_defect_2019} and molecules. In these cases the screening lengths and resulting optimal parameters within the HSE functional are often quite different from bulk. Unfortunately these are exactly the conditions in our current system, and it is not \textit{a priori} clear how the mixing and screening parameter can be properly handled. Finding the appropriate computational parameters for different systems in this study in order to yield consistency and exactitude is far from trivial and beyond the scope of this work.\\
The GGA-PBE functional is sufficient for a qualitative exploration of this problem. It remains a highly used and versatile method, successfully applied to study point defects in bulk, layered and slab systems: from the object of this study \NV\, to metal impurities \cite{prasad_charge_2023,goss_density_2024,smeltzer_13c_2011,gali_identification_2009}, different diamond orientations and surfaces, again with \NV\, inside \cite{reed_diamond_2022,rivero_surface_2016,meara_computational_2020}. Related methods were previously successfully applied to determine molecular levels, hyperfine structure and geometry of \aza\, radical molecules but using LDA\cite{stergiou_long-lived_2019,rio_electronic_2018,Long_tanuma_2023,Robust_tanuma_robust_2021}. While keeping in mind the gap underestimation and valid criticisms of GGA, it still nonetheless permits a broad range of study, geometric optimisation, electronic band structure, hyperfine coupling\cite{smeltzer_13c_2011,gali_identification_2009}, vibrational calculation\cite{goss_identification_2014}, formation energies while remaining consistent between those properties. For example, Neethirajan \textit{et al} used this methodology to successfully study diamond hydrogenated surface with \NV\, inside, and the impact of water molecules on the NV charge under band bending considerations \cite{neethirajan_controlled_2023}.

\begin{table}[!t]
    \centering
    \begin{tabular}{ l c c c } 
        \hline\hline
        Systems & $\omega_e$ & $\omega_a$ & E$_{x-y}$  \\
        \hline\hline
        2$\times$2$\times$2 & 760 & 233 & 387 (300) \\
        
        3$\times$3$\times$3 & 86 & 146 & 86 (50) \\
        
        4$\times$4$\times$4 & 18 & 33 & 14 (8) \\
        \hline\hline
    \end{tabular}
    \caption{Dispersion for \aun\, ($\omega_a$), $e$ ($\omega_e$) states and the splitting of the $e$ states (E$_{x-y}$) for \NV\, in different $n\times n \times n$ diamond supercells (meV). 
    Bracketed literature values are calculated using LDA with large wave-function basis set for the carbon \cite{H.Pinto_bandst_PhysRevB.86.045313}, rather than a smaller set used in our calculations.}
    \label{tab:tableau.1}
\end{table}

\section{\label{sec:level3}Results}

The computational model is initially benchmarked using an \NV\, defect in bulk diamond, before creating 1~nm thick (001) hydrogenated and fluorinated diamond slabs with lateral cell dimension choice based on the bulk system results.
The electron affinity of the constructed surfaces is evaluated as well as the spin population over the system eigenstates through Mulliken analysis. Finally, the spin active \azar\,  is added to the surface, evaluating its stability and interaction with the \NV\,  centre.
Calculated slab  band gap, eigenstates energies of all \NV\,  electronic states for bulk diamond, H- and F- terminated surfaces and heterogenous systems (with added \aza) are summarised in Table \ref{tab:tableau.2}, with full details given in the Table.S4 of the S.I.

\subsection{\NV\, in bulk diamond}

In all cases, we find \NV\, defect with $S=1$ in bulk diamond as the most stable configuration, as shown in Fig. \ref{fig:figure.supercell} where the 4$\times$4$\times$4 supercell (512 atoms) electronic band structure is shown together with associated NV molecular orbitals.

\begin{figure}[!t]
\includegraphics[scale=0.9]{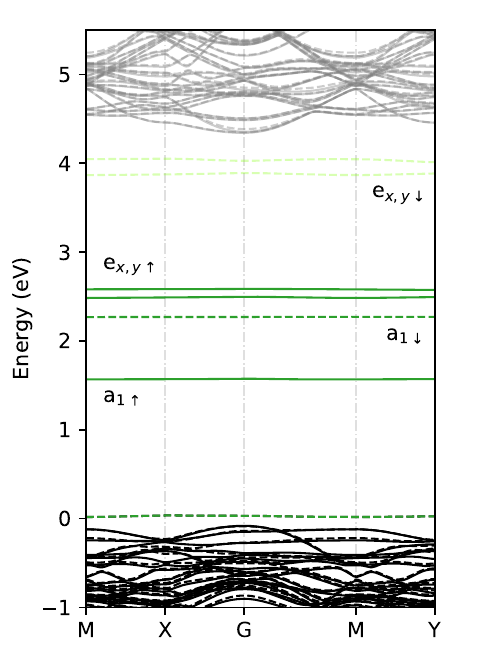}
\caption{Electronic band structure of (2$\times$1)-(001) hydrogenated diamond surface with a single \NV\, embedded in the centre of the diamond slab. Zero is set to the top of the valence band. Solid/dashed lines represent spin-up/-down electrons respectively, while dark/faded lines showing filled/empty states respectively. Black and green colours indicate diamond slab  and NV defect related eigen states respectively.}
\label{fig:figure.hydro.NV-}
\end{figure}

\begin{figure}[!t]
\includegraphics[scale=0.9]{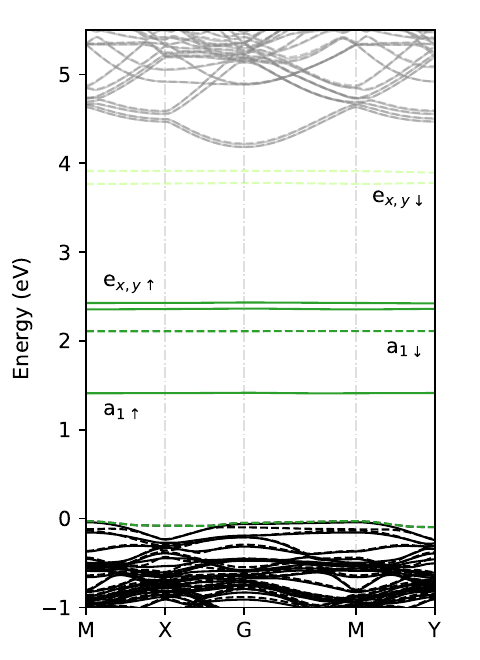}
\caption{Electronic band structure of (2$\times$1)-(001) fluorinated diamond surface with a single \NV\,  embedded in the centre of the diamond slab. Zero is set to the top of the valence band. Solid/dashed lines represent spin-up/-down electrons, respectively, while dark/faded lines showing filled/empty states, respectively. Black and green colours indicate diamond slab  and NV defect related eigen states, respectively.}
\label{fig:figure.fluor.NV-}
\end{figure}

The maximum observed lifting of the $e$ doublet (in both spin up and down) state degeneracy at a specific k-point,  E$_{x-y}=14$~meV, and the sizable dispersion of these electronic bands, $\omega(k)$ (Table \ref{tab:tableau.1}), emerge due to the  
coupling between the \NV\, defects in neighbouring cells.

Typically interested in commercially available nanodiamonds in the ppm range, in this study we are well above that value ($\sim$2000ppm). Additionally, those diamonds contain additional anisotropy sources that affect \NV\, such as vacancies, substitutional atoms or surface strains. Modifying the defect density can significantly alter E$_{x-y}$ and $\omega$. \\

Next, we study the impact of the defect-defect distance, i.e., the NV density, on the dispersion $\omega$ of the $a$ and $e$ states, where $\omega_e$ is the largest dispersion between \ex\,  and \ey\,  (Table \ref{tab:tableau.1}).
As expected, we observe a decrease of both the dispersion and the splitting as the system size and defect spacing increase. Notably, the band splitting E$_{x-y}$ follows closely the trends obtained with a larger wave function basis set type of calculation\cite{H.Pinto_bandst_PhysRevB.86.045313}, with only a negligible 6~meV difference between our calculations and the literature, for the largest comparable supercell (512 atoms). This agreement validates our choice of smaller carbon basis set to describe the diamond by reducing the computational cost while conserving the same accuracy. 
We note that in all bulk supercells the degeneracy is reduced to $0$ in the $\Gamma$ symmetry point. \\

\begin{figure*}[!t]
\includegraphics[scale=0.72]{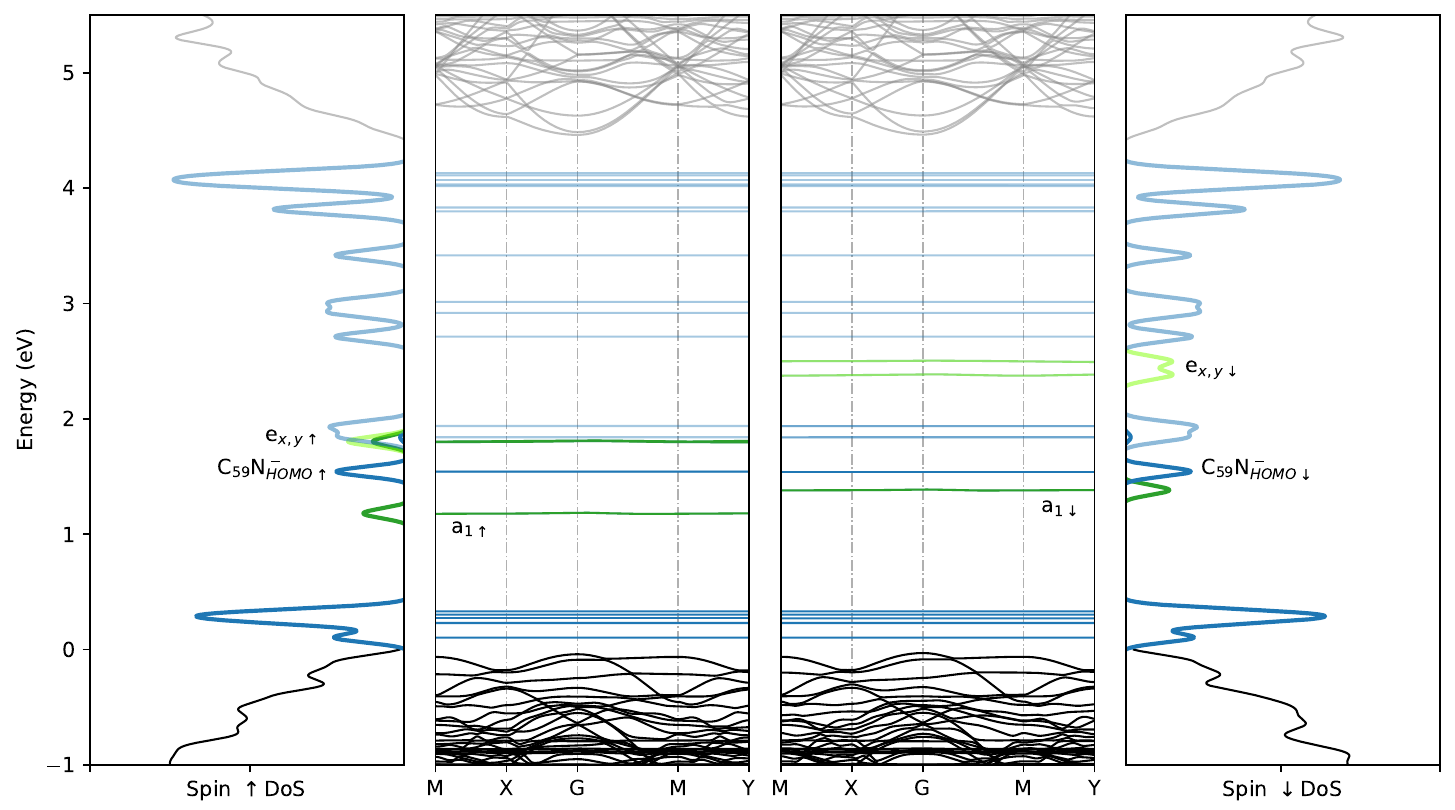}
\caption{Electronic band structure and projected density of states (pDoS) of \azam\, on top of (2$\times$1)-(001) hydrogenated diamond surface system with a single neutral NV defect (NV$^0$). Left and right half side correspond to the spin up- and down-states respectively. Zero is set to the top of the diamond slab\, valence band. Dark/faded lines show filled/empty states respectively. Black, green and blue colours indicate slab\, diamond, NV defect and \aza$^0$ related eigen states respectively, which is identified by the pDoS.}
\label{fig:figure.hydro.bandstxdos}
\end{figure*}

Fig. \ref{fig:figure.supercell} shows the molecular orbital of each defect state, where the influence of the electronic bands dispersion is already included.
The $a_1$ state shows its expected orbital composition, made of both the N lone-pair and the C-$sp^3$ orbitals.
The $e$ state should have purely C-sp$^3$ orbital character  with $e_x$ composed of two carbon orbitals, and $e_y$ of three \cite{doherty_negatively_2011,PhysRevLett.101.226403}.
This description is also verified in our calculation when integrating the Kohn-Sham functions at $\Gamma$ point. The slight asymmetry visible in the $e_x$ state is due to dispersion caused by the finite cell size.
The asymmetry is further exacerbated by the strain effects due to the defect-defect interactions.

\subsection{\NV\,  and (2$\times$1)-(001) hydrogenated diamond surface}

Keeping the same defect-defect distance from the 4$\times$4$\times$4 bulk supercell (xy plane), we next construct a hydrogenated (2$\times$1)-(001) diamond surface. The shallow defect is placed in the middle of the slab at a distance of 0.45~nm\, from the surface. Mulliken analysis demonstrates that the negative charge and unpaired electron population localise on the nitrogen and the vacancy neighbours ($>$90\%), respectively, similar to the bulk case discussed above.
Fig. \ref{fig:figure.hydro.NV-} shows an E$_{x-y}$ splitting increase to a maximum of 97~meV for the up and 146~meV\, for the down configurations. We note that the dispersion does not change, showing it comes solely from the defect density. The $e$ states degeneracy is of a similar order of magnitude to the bulk 3$\times$3$\times$3 supercell.
The surface strain induces an additional degeneracy lift of the $e$ doublet over the entire Brillouin zone, even at the $\Gamma$ point, which is in contrast to the impurity density related strain present and discussed for the bulk supercell.

To clarify the discussion we define a slab band gap which describes the band gap of the slab system or hetero-system. The difference between the bulk and slab band gap lies in the appearance of surface states when modeling diamond slabs. Studying these states are important and extended studies of hydrogenated and fluorinated slab surface states as been done using a different methodology \cite{chou_nitrogen-terminated_2017,chou_nitrogen-vacancy_2017}. Following the discussion in Section.\ref{fucntional}, we do not make a detailed quantitative analysis of our results here, focusing instead on describing the electronic stability of our systems. 

Hydrogenation of the surface does not introduce surface states deep into the slab band gap, as seen in other study (Rydberg states), which can be understood through our use of localised Gaussian based orbitals rather than plane waves. Nonetheless Mulliken analysis and pDoS on the first H layer confirms the presence of surface states at the bottom of the conduction band, consistent with the literature (see Fig.S5).  In our case they are not overlapping with the high-lying \NV electronic states.

\begin{figure*}
\includegraphics[scale=0.72]{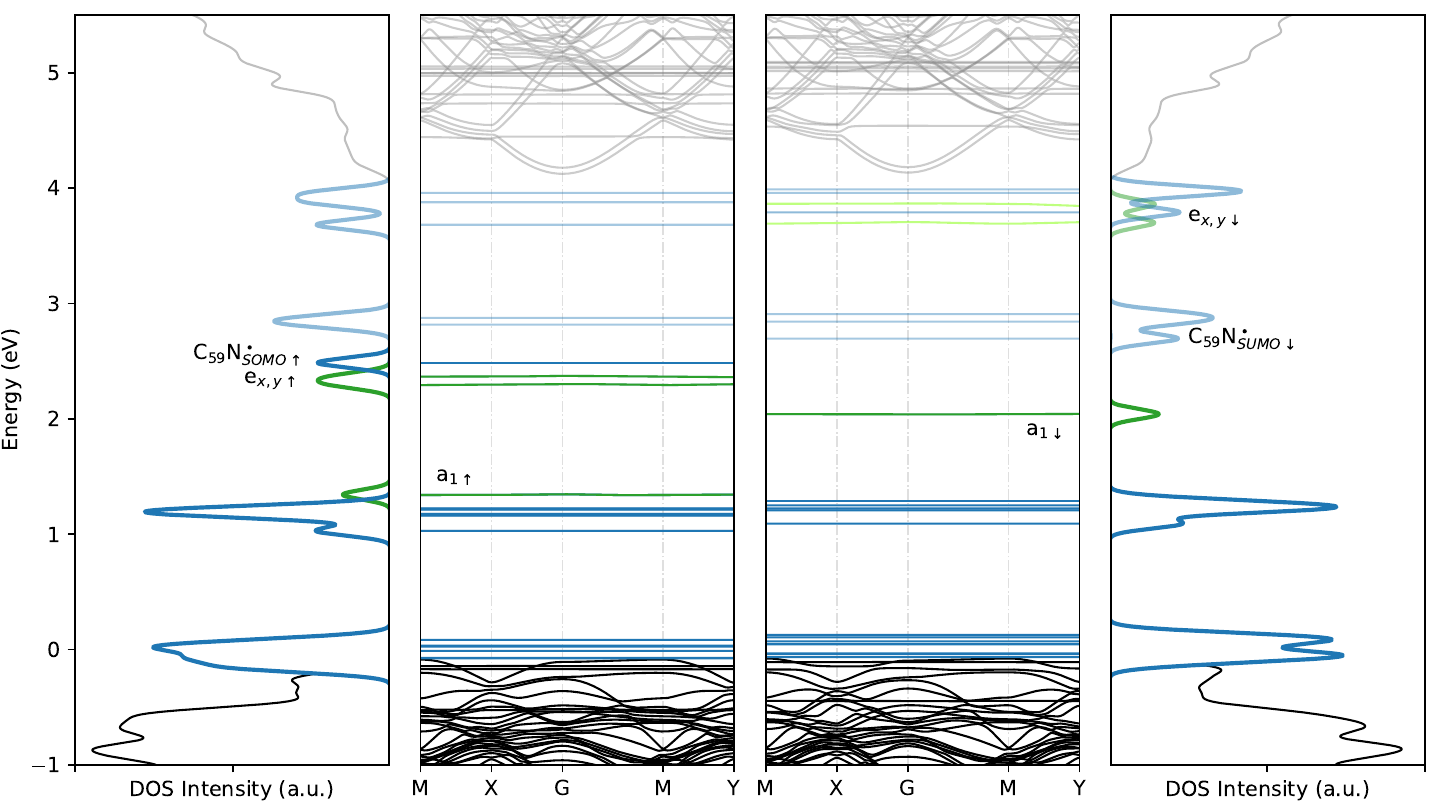}
\caption{Electronic band structure and the projected density of states (pDoS) of \aza on top of (2$\times$1)-(001) fluorinated diamond surface system with a single negatively charged \NV\, defect embedded inside. Left and right half side correspond to the spin up- and down-states respectively. Zero is set to the top of the diamond slab\,valence band. Dark/faded lines show filled/empty states respectively. Black, green and blue colours indicate slab\, diamond, NV defect and \aza$^0$ related eigen states respectively, which is identified by the pDoS}
\label{fig:figure.fluor.bandstxdos}
\end{figure*}

The proximity to the hydrogenated surface shifts the \NV\,  defect levels up in the gap by 0.5-0.6~eV (zero of energy set at the top of the valence band).
Notably, while the $a_1'$ down state lies deep inside the valence band in bulk diamond, it emerges from the diamond valence band in this case.\\
The standard definition of the electronic affinity, EA=E$_{\rm vacuum}$-E$_{\rm CBM}$, where E$_{\rm vacuum}$ is the vacuum energy of the system and E$_{\rm CBM}$ is the conduction band minimum. Since GGA-PBE underestimates band gap values, for the hydrogenated slab we multiply E$_{\rm CBM}$ by 1.23, scaling the gap to the experimental diamond band gap, E$_g$=5.46~eV. This gives a negative electron affinity (NEA) for the hydrogenated slab of -1.66~eV, which agrees well with literature  \cite{Sque_structure_2006,Tiwari_PhysRevB.84.245305,Robertson_robertson_band_1998,kaviani_proper_2014}.
We note that the slab band gap is 0.14~eV higher than the bulk value, which is another demonstration of the importance of hydrogenated surfaces for shallow \NV\, centres in nanodiamonds. The upward shifts of the \NV\,  defect states as well as the increased band gap are closely linked to the NEA. This may have important consequences for potential charge transfer behaviour at the surface when it is in contact with an adsorbed species. The \NV\, proximity to the (2$\times$1)-(001) hydrogenated surface influences the degeneracy but leaves the \NV electronic configuration itself intact (Fig. \ref{fig:figure.hydro.NV-}). We note that these variations have to be moderated by the quantum confinement imposed by these finite slabs.

\subsection{\NV and (2$\times$1)-(001) fluorinated diamond surface}

We next examine a comparable (2$\times$1)-(001) diamond slab, but this time with a fluorinated surface. The \NV\, defect is positioned at  $\sim 0.48$~nm depth. The excess charge and spin population localises mostly on the \NV\, defect ($\sim$90\%), similarly to both the hydrogenated surface and bulk diamond cases. This  demonstrates that the shallow \NV\, defect survives as a spin active \NV\, centre when the surface is fluorinated.Changes in the electronic band structure (Fig. \ref{fig:figure.fluor.NV-}) show either weaker or opposing trends to those of the hydrogenated surface.

In this case, the slab band gap only slightly decreases by 0.03~eV compared to bulk diamond, related to the absence of surface states in the conduction band giving instead looping states inside the slab. Unlike the H-terminated surface, surface states can be found in the valence band as seen in Fig.S4.

All \NV\, defect-related states shift upward in energy compared to the bulk, but the effects are less pronounced than with the hydrogenated surface. Specifically, the shifts are now in the range of 0.3 to 0.4~eV.

The degeneracy lifting between the $e$ up- levels is comparable, as $E_{x-y}$ decreases from 97~meV to 72~meV. 
Applying the same scaling method as for the hydrogenated case, this time with a scaling factor of 1.30 the fluorination makes the surface electron affinity positive (PEA) at about +1.89~eV, which is smaller than expected from the literature data but still can be considered as a high PEA \cite{Tiwari_PhysRevB.84.245305, kaviani_proper_2014}.

In summary, shallow \NV\,  beneath a fluorinated surface shows the same reconstruction and similar surface proximity effects to the hydrogenated surface case.
However the system presents a PEA, in direct opposition to the hydrogenated surface case.

We see in the next section that this important difference between the  surface functionalities profoundly affects the behaviour of adsorbed spin-active radical, \azar, with both NEA and PEA surface systems sheltering a shallow NV$^-$ defect.

\subsection{Spin active molecule on (001) diamond surfaces}

We next bring the radical azafullerene into the proximity of diamond surfaces. Fullerenes are known as very good electron acceptors. The spin active fullerene, while charge neutral, has a singly-occupied molecular orbital (SOMO) making \azar\, a radical species, which can potentially trap an additional charge \cite{Long_tanuma_2023,Robust_tanuma_robust_2021}. Both the radical and the extended $\pi$-states on the cage are highly sensitive to environmental perturbation, including the proximity of a negatively charged NV inside the slab and the EA at play on the surface. All these factors may compromise the computational convergence, when starting with the neutral atom electronic distribution before redistributing the electrons during the self-consistency cycle. We report here the most stable electronic configuration for the two EA cases, namely the adsorbent radical \azar\, on either H- and F-terminated (001) diamond surfaces with an underlying subsurface \NV\, centre.

\subsubsection{\aza\, on a (2$\times$1)-(001) H-terminated diamond surface with a sub-surface \NV defect}

We first investigate the hydrogenated surface with a \aza\, adsorbed on the surface just above the sub-surface NV defect.
After structural and electronic relaxation, Mulliken analysis of the charge, spin and eigen states in this combined system demonstrate a negatively charged azafullerene \azam, and a neutral nitrogen vacancy, NV$^0$.  
Compared to the initially isolated stable \azar\,  and \NV, the final state suggests a one-electron charge transfer to the fullerene cage. 
In this new charge state, \azam\, loses its spin, and the magnetic field sensing capability of the NV centre is completely quenched.

The electronic band structures and the pDoS for both up- and down-spins are shown in Fig.\ref{fig:figure.hydro.bandstxdos}.  Due to the large number of gap states, the pDoS is also included as a guide to the eye.
The calculated suface band gap\, increases by 116~meV\, when compared to the bare structure with a \NV\, beneath a clean H-terminated surface, reaching 4.46~eV. For comparison, we find the slab band gap getting closer to a hydrogenated slab with sub-surface NV$^0$ without \aza\,  present (see Supporting Information). Therefore, the increase in slab band gap is mainly due to charge transfer that changes the charge state of \NV\, to NV$^0$ and affects the slab band gap at the same time by moving the surface states in the conduction band. 

Interestingly, the splitting of the $e$ doublet for the up and down configuration are not equivalent; without \aza\,  present, E$_{x-y}$ is 97 and 146~meV for spin-up and -down, but when the azafullerene is in the vicinity of the defect $e_\uparrow$ states (populated by 1 electron) are split by only 13~meV, while the empty $e_\downarrow$ states are separated by 172~meV.

Due to the charge transfer associated with the adsorption of \aza, the former SOMO of the \azar\, becomes a fully occupied HOMO of \azam, with both the up- and down-states aligned in energy. This state is lying lower in energy than the highest occupied $e$ state of the NV defect. These $e$ up-states are also nearly degenerate with the first LUMO \azam\, states, raising the intriguing possibility that the small local electronic excitations could give rise to further charge transfer from the NV centre to the \aza, creating NV$^+$ centres.  Indeed, when we run the same calculation starting from neutral NV$^0$ the final stable configuration is NV$^+$ and \azam.

\subsubsection{\aza\, on a (2$\times$1)-(001) F-terminated diamond surface with a sub-surface \NV defect}

\begin{table}[!b]
\centering
\scalebox{0.89}{
\begin{tabular}{l c c }
    \hline\hline
     System & E$_g$ & E$_{x-y}$\\
     \hline\hline
     Bulk Diamond & 4.210 & 14(14)  \\
     \hline
     NV$^-$ H-terminated & 4.347 & 97(146) \\
     NV$^-$ F-terminated & 4.184& 72(147)  \\
     \hline
     \azam $\perp$ NV$^0$ H-terminated & 4.463 & 13(125) \\
     \azar $\perp$ NV$^-$ F-terminated & 4.134 & 74(172) \\
    \hline\hline
\end{tabular}}
\caption{Calculated band gap (bulk) and slab band gaps E$_g$ (eV), the maximum $e_{\uparrow}$ ($e_{\downarrow}$) splitting (meV) for bulk diamond as reference, followed by the \NV\ with hydrogenated and fluorinated surfaces, and the combined system with \azar\, molecule on the slab surface.}
\label{tab:tableau.2}
\end{table}

As discussed above, surface passivation by fluorination gives PEA character to the surface. Adding a \aza\, to the surface thus results in very different behaviour compared to the hydrogenated surface. In the fluorinated case the charge, spin and eigen state population show that the most stable electronic distribution involves a neutral radical \azar, and a negatively charged NV centre, \NV\, i.e. both species maintain their electronic spin and charge states in the hetero-system.
Fig. \ref{fig:figure.fluor.bandstxdos} shows the electronic band structure and the pDoS for the case of fluorinated surface.  In the presence of \azar, the diamond slab band gap decreases further (50~meV)\, relatively to the pristine F-terminated surface and the bulk system, similarly the populated \NV\,  defect states energetically shift down (by 50-70~meV) relatively to their positions without the presence of azafullerene.

The azafullerene SOMO contains a single up-electron, matching the isolated neutral species (Fig. \ref{fig:figure.aza}).
The pDoS shows that the $a_{1\uparrow}$ state lies above the fullerene $h_u$ states, which slightly splits due to the symmetry lowering.
There is an upward shift of all the \azar\,  levels compared to the hydrogenated surface system, allowing identification of the symetrically broken azafullerene  g$_g$ and h$_g$ states (in total 9 eigenstates) just above the diamond valence band.

We note that both the conduction and valence bands demonstrate crossing and anti-crossing with the \azar\,  states. However their position in the bulk electronic structure makes precise assignment difficult.

\section{Discussion and summary}

In this study we have explored shallow \NV centres beneath both hydrogenated and fluorinated (001) diamond surfaces using density functional methods, both with- and without surface adsorbed azafullerene \aza\, species.

The hydrogenated (001) diamond surface has a calculated NEA of -1.66~eV. Such a strongly negative value is known to facilitate band bending at the diamond surface region, with potential to aid charge transfer between an adsorbent and the surface \cite{Tiwari_PhysRevB.84.245305,Robertson_robertson_band_1998,kaviani_proper_2014}. 
In contrast, the fluorinated surface has a calculated\, positive electron affinity (PEA) of +1.89~eV, which shows the least modifying behaviour of the \NV, bulk diamond electronic band structure.

Positioning \NV\, in proximity to either surface increases the splitting between its occupied $e_\uparrow$ or unoccupied $e_\downarrow$  states.
This splitting is expected to broaden the photoluminescence of such shallow defects and could be experimentally detectable \cite{H.Pinto_bandst_PhysRevB.86.045313, Ofori_PhysRevB.86.081406}. 
We note that the calculations show the energetic depth of the \NV\,  states in the gap is linked to the physical depth of the defect centre below the surface, which may provide a useful spectroscopic probe for \NV\, centre depth profiling.

Adding the neutral spin active molecule \azar\, to these systems provides insight into the charge doping effect.
In the case of the hydrogenated surface the most stable electronic configuration is a negatively charged azafullerene, \azam, on the surface with a sub-surface neutral NV centre, NV$^0$, beneath it.
This is consistent with a charge doping effect induced by the NEA, where the negative charge of the defect can easily migrate to the surface due to its electron affinity, which is afterwards captured by the acceptor-like behaviour of the fullerene. 
Our results suggest that even very small electronic excitations may result in even further charge redistribution and possible formation of NV$^+$. However, our ground state calculations are not capable to reveal the details of such transient electron dynamics.  

We conclude that azafullerene in the vicinity of a hydrogenated (001) diamond surface with sub-surface \NV\,  will undergo charge transfer from the defect to the \aza, killing its radical state and spin activity.  Such charge transfer effects may not occur when the \NV\,  centre is sufficiently deep beneath the surface. Although computational resources preclude a complete investigation of NV$^-$ stability as a function of depth, single point energy calculations of a thicker slab showed that even at a depth of 1.1~nm, the \NV, defect still loses one electron to the \aza.  This is consistent with quenching of \NV\ photoluminescence we observe experimentally when mixing \aza\, with hydrogenated nanodiamonds, with a mean volumetric size of 40~nm. 

We note that the EA for hydrogen- and fluorine- terminated surfaces is largely independent of surface orientation and reconstruction ($\pm$1 eV) \cite{Tiwari_PhysRevB.84.245305,Robertson_robertson_band_1998,kaviani_proper_2014,DIEDERICH_1998219,Bandis_PhysRevB.52.12056,BANDIS_1996315,Rietwyk_10.1063/1.4793999,Humphreys_10.1063/1.118545,Rutter_PhysRevB.57.9241,BAUMANN_1998320,Cui_PhysRevLett.81.429,Kern_PhysRevB.56.4203,LI_2019273}, showing the effects calculated here should equally apply to other surface orientations and diamond nanoparticles.

In contrast, surface fluorination may provide a useful route to block unwanted charge transfer processes. When passivated with fluroine, sub-surface NV maintains its negative charge state in the presence of surface azafullerene, which remains a neutral radical, \azar. This conserves both the spin activity of the molecule and the important photoluminescence properties of the defect.
We associate this stability to the down band bending effect from the PEA of the F-terminated surface, that prevents the additional electron on the NV transferring to the adsorbent.
To investigate the spin dynamics of the \azar\, or  other similar spin-active molecules through the photoluminescence of the shallow \NV\, defects, fluorination of the (nano)diamond surface may be a key first step.

\begin{acknowledgments}
This work received financial support from the French government under the EUR LUMOMAT project and the Investments for the Future program ANR-18-EURE-0012.
The research was supported by the Slovenian Research agency under project grant PR-13384, the research program P1-0125 and the research project J1-3007.
We acknowledge COST Action CA21126 'Carbon molecular nanostructures in space (NanoSpace)', supported by COST (European Cooperation in Science and Technology).
All DFT calculations were performed at the CCIPL (Centre de Calcul Intensif des Pays de la Loire).
\end{acknowledgments}

\clearpage
\bibliographystyle{apsrev4-1}
\bibliography{article}

\end{document}